\documentclass[journal=jpccck,manuscript=article]{achemso}
\SectionNumbersOn
\usepackage{chemformula} 
\usepackage{comment}
\usepackage[T1]{fontenc} 
\usepackage{braket}
\usepackage{ulem}
\usepackage{float}
\usepackage{xr}
\newcommand*\mycommand[1]{\texttt{\emph{#1}}}
\DeclareMathAlphabet{\altmathcal}{OMS}{cmsy}{m}{n}
\DeclareMathOperator{\Tr}{Tr}
\newcommand{\eq}[1]{Eq.~\eqref{#1}}

\newcommand{\xccomment}[1]{}
\newcommand{\note}[1]{{\bf[{\color{brown}#1}]}} 
 
\renewcommand{\note}[1]{#1}
\renewcommand{\sout}[1]{}
\author{Roman Korol}
	\email{roman.korol@rochester.edu}
	\affiliation{Department of Chemistry, University of Rochester, Rochester, New York 14627, USA}
\author{Xinxian Chen}
	\affiliation{Department of Chemistry, University of Rochester, Rochester, New York 14627, USA}
\author{Ignacio Franco}
	\email{ignacio.franco@rochester.edu}
	\affiliation{Department of Chemistry, University of Rochester, Rochester, New York 14627, USA}
	\affiliation{Department of Physics, University of Rochester, Rochester, New York 14627, USA}

\title{High-frequency tails in spectral densities}

\begin{document}



\begin{abstract}
    Recent advances in numerically exact quantum dynamics methods have brought the dream of accurately modeling the dynamics of chemically complex open systems within reach. 
    Path-integral-based methods, hierarchical equations of motion (HEOM) and quantum analog simulators all require the spectral density (SD) of the environment to describe its effect on the system. 
    Here we focus on the decoherence dynamics of electronically excited species in solution in the common case where nonradiative electronic relaxation dominates and is much slower than electronic dephasing. 
    We show that the computed relaxation rate is highly sensitive to the choice of SD representation –– such as the Drude-Lorentz or Brownian modes -- or strategy used to capture the main SD features, even when early-times dephasing dynamics remains robust.
    The key reason is that \note{electronic} relaxation is dominated by the \note{resonant contribution from the} high-frequency tails of the SD, which are orders of magnitude weaker than the main features \note{of the SD} and can vary significantly between strategies. 
    This finding highlights an important, yet overlooked, numerical challenge: obtaining an accurate spectral density requires capturing its structure over several orders of magnitude to ensure correct decoherence dynamics at both early and late times. To address this, we provide a simple transformation that recovers the correct relaxation rates in quantum simulations constrained by algorithmic or physical limitations on the shape of the SD.
    Our findings enable comparison of different numerically exact simulation methods and expand the capabilities of analog simulations of open quantum dynamics. 
\end{abstract}

\section{\label{intro}Introduction}
Accurate calculations of the quantum dynamics of open quantum systems with chemically complex environments would advance our understanding of many problems of interest in chemistry, biology, and quantum information science. Tuning of the coherence times of molecular qubits,\cite{zadroznyMillisecondCoherenceTime2015} learning from efficient energy transfer in photosynthesis,\cite{caoQuantumBiologyRevisited2020} and \textit{in silico} design of molecular engines\cite{aprahamianFutureMolecularMachines2020} - these are just select examples of the many new possibilities that would open up.
Due to the inherent complexity of the open quantum dynamics, there isn't a single universal approach to solve it. Instead, a great variety of methods has been developed, each of them with a different regime of applicability. 

The main challenge in open quantum dynamics is to accurately describe the effect of a large environment on a small system of interest. The unitary dynamics of system plus environment is intractable owing to the large (possibly macroscopic) environment.  Historically, two main strategies to make the problem tractable without the introduction of uncontrolled approximations have emerged;
we will refer to these two strategies as "unitary" and "reduced" for brevity. In the unitary approach, finite number of degrees of freedom of the environment are included and the total dynamics (of both system and truncated environment) is unitary. Multiconfigurational time-dependent Hartree (MCTDH),\cite{meyerMulticonfigurationalTimedependentHartree1990,worthUsingMCTDHWavepacket2008} and its multilayer extension,\cite{wangMultilayerFormulationMulticonfiguration2003} the density matrix renormalization group (DMRG),\cite{whiteRealTimeEvolutionUsing2004} the time-evolving density matrix using the orthogonal polynomials algorithm (TEDOPA),\cite{chinExactMappingSystemreservoir2010} the time-dependent Davydov ansatz,\cite{zhouFastAccurateSimulation2016} and the effective-mode (EM) approach\cite{cederbaumShortTimeDynamicsConical2005} all utilize this strategy. Only a finite number of environment modes can be included, meaning that the overall dynamics is reversible. Thus, the system cannot reach the thermal state even in principle and instead will experience recurrences in long-time dynamics. On the other hand, these methods are applicable to "tough" cases, such as anharmonic environment modes,\cite{wangQuantumDynamicalSimulation2007} failure of the Born-Oppenheimer approximation,\cite{huLessonsElectronicDecoherence2018} or strong,\cite{priorEfficientSimulationStrong2010} possibly nonlinear\cite{worthUsingMCTDHWavepacket2008} coupling to the environment. 

Hierarchical equations of motion (HEOM)\cite{tanimuraNonperturbativeExpansionMethod1990} and its variants\cite{ishizakiUnifiedTreatmentQuantum2009,tangExtendedHierarchyEquation2015,chenBexcitonicsQuasiparticleApproach2024} as well as real-time path integral (PI) methods\cite{makriTensorPropagatorIterative1995,makriBlipDecompositionPath2014,strathearnEfficientNonMarkovianQuantum2018,makriSmallMatrixPath2021,lambertQuantumclassicalPathIntegral2012,lambertQuantumclassicalPathIntegral2012a,kunduRealTimePathIntegral2020} utilize the reduced strategy to make open quantum dynamics numerically tractable. "Reduced" here refers to the fact that only system dynamics is followed explicitly; the effect of the environment is captured implicitly by introducing a bath of  large (possibly uncountably infinite) number of harmonic degrees of freedom, each bilinearly coupled to the system. The effect of this harmonic bath on the system dynamics on one hand can be captured exactly within an effective description,\cite{feynmanTheoryGeneralQuantum1963,leggettDynamicsDissipativeTwostate1987} and on the other hand it can mimic\cite{makriLinearResponseApproximation1999,chandlerIntroductionModernStatistical1987} the effect of a complex environment through Gaussian response. 
For example, recently, the relaxation in bacterial light harvesting chromophore has been simulated\cite{kunduB800toB850RelaxationExcitation2022} using the small matrix decomposition of the path integral algorithm.\cite{makriSmallMatrixPath2021}  Similarly, HEOM was used to model the FMO complex.\cite{lambertQuTiPBoFiNBosonicFermionic2023}

Recent advances at the intersection of quantum information science and theoretical chemistry opened up a radically new set of approaches to obtain accurate open quantum dynamics -- using digital quantum computation\cite{huQuantumAlgorithmEvolving2020,wangSimulatingOpenQuantum2023} or analog quantum simulation\note{, where the dynamics of interest is mapped onto a highly controllable experimental system}.\cite{macdonellAnalogQuantumSimulation2021,kimAnalogQuantumSimulation2022,mostameEmulationComplexOpen2016,daleyPracticalQuantumAdvantage2022} The former has many of the strengths and weaknesses of the unitary methods, since quantum computers are naturally suited to describe unitary quantum dynamics. In contrast, quantum analog simulators can be used to realize open system dynamics with macroscopically large environments,\cite{kimAnalogQuantumSimulation2022,mostameEmulationComplexOpen2016} which groups them together with the reduced methods.

Both the PI and HEOM methods are numerically exact, meaning the dynamics of the system can be described as accurately as needed by tightening convergence parameters appropriately. 
Assuming the dynamics is converged, the model we formulate to describe a process is the only remaining source of discrepancies between a numerically exact simulation and physical reality.  In this paper, we analyze the importance of faithful representation of a structured bath, illustrating our findings using HEOM simulations.\cite{chenBexcitonicsQuasiparticleApproach2024} However, we emphasize that our findings extend to any reduced approach (all HEOM variants, all real-time PI methods, and quantum analog simulators \note{of open quantum dynamics}\sout{ with continuous baths}).

The properties of any harmonic bath bilinearly coupled to the system are fully captured\cite{feynmanTheoryGeneralQuantum1963} by the bath spectral density (SD) defined in the frequency domain as
\begin{equation}\label{eq:J}
    J(\omega)=\sum_j\hbar|g_j|^2\left(\delta(\omega-\omega_j) - \delta(\omega+\omega_j)\right)
\end{equation}
with each bath mode $\omega_j$ characterized by the system-bath coupling constant $g_j$ and $\hbar$ denoting the reduced Planck's constant. The spectral density enters the equations of motion of the open quantum system via the bath correlation function (BCF) also known as bath response function,\cite{feynmanTheoryGeneralQuantum1963}
\begin{equation}\label{eq:bcf}
    C(t)=\int_{-\infty}^\infty d\omega J(\omega)\left(1+n(\beta\omega)\right)e^{-i\omega t},
\end{equation}
where as usual $\beta=(k_BT)^{-1}$ is the inverse temperature multiplied by the Boltzmann constant, and $n$ denotes the Bose-Einstein distribution $n(x)=(e^{x}-1)^{-1}$.

Obtaining an accurate SD for a system that interacts with a structured environment is highly non-trivial. Molecular dynamics (MD) or hybrid quantum mechanics/molecular mechanics (QM/MM) methods can be used to calculate classical (i.e., real only) BCF and the SD can be constructed from it. \cite{olbrichTheorySimulationEnvironmental2011,damjanovicExcitonsPhotosyntheticLightharvesting2002,shimAtomisticStudyLongLived2012,valleauAlternativesBathCorrelators2012,wiethornCondonLimitCondensed2023} Experimental data, such as fluorescence line narrowing spectra\cite{rengerRelationProteinDynamics2002,valleauAlternativesBathCorrelators2012} or resonance Raman scattering spectra,\cite{gustinMappingElectronicDecoherence2023} can also inform the construction of SD for a pure dephasing process. However, both experimental and theoretical approaches have a limited regime of validity and imperfect accuracy.\cite{maityMultiscaleQMMM2021} 
The low frequency end of the spectrum (that is, $\lim_{\omega\rightarrow0^+}J(\omega)\propto\omega^s$) is known to sensitively affect the dynamics,\cite{leggettDynamicsDissipativeTwostate1987} leading to qualitatively distinct features in the ohmic($s=1$), subohmic ($s<1$) and superohmic ($s>1$) cases. In contrast, the details of the high-frequency end of the spectrum are believed to be largely inconsequential with the notable exception of the polaronic dressing of charge carriers.\cite{bundgaard-nielsenNonmarkovianPerturbationTheories2021} 

 Indeed, it is common to see simulations employing different high-frequency cut-off functions\cite{tanimuraReducedHierarchyEquations2012,kellShapePhononSpectral2013,liuReducedQuantumDynamics2014,ritschelAnalyticRepresentationsBath2014} decaying as fast as exponential\cite{kunduB800toB850RelaxationExcitation2022} ($\lim_{\omega\rightarrow\infty}J(\omega)\propto e^{-\omega/\omega_c}$) or as slow as inverse polynomial ($\lim_{\omega\rightarrow\infty}J(\omega)\propto \omega^{-p}$) with $p=1$ (Lorentzian),\cite{maityMultiscaleQMMM2021,boseMultisiteDecompositionTensor2022} $p=3$ (Brownian),\cite{knoxLowtemperatureZeroPhonon2002} and $p>3$.\cite{ritschelAnalyticRepresentationsBath2014} The precise form of the high-frequency cutoff is motivated by the exact results in limiting cases (for example the underdamped Brownian peaks),\cite{tanimuraReducedHierarchyEquations2012,liuReducedQuantumDynamics2014,ritschelAnalyticRepresentationsBath2014} or by numerical efficiency considerations. In fact, HEOM simulations are frequently performed with the Drude-Lorentz form of the SD cutoff, as it is particularly efficient;\cite{liuReducedQuantumDynamics2014,ritschelAnalyticRepresentationsBath2014} oppositely, the exponential form of the cutoff function is preferred for MCTDH\cite{meyerRegularizingMCTDHEquations2018} and other methods that require discretization of the SD and, therefore, have a maximum frequency. Physical constraints can similarly limit the choice of the peak functional form in quantum analog simulators.\cite{kimAnalogQuantumSimulation2022} It is widely assumed that the precise choice of the cutoff is irrelevant and only the total contribution of the high-frequency tails to the reorganization energy should be considered to recover the correct dynamics.

In this study, we reexamine this assumption with a focus on the dynamics of coherent electronic excitation in condensed phase. The decoherence dynamics typically proceeds in two steps: initial fast loss of phase information, followed by the much slower population relaxation. We aim to accurately capture both the fast and the slow components of the decoherence dynamics. We focus on the HEOM method, requiring that the BCF is written as a finite sum of exponentials,\cite{chenBexcitonicsQuasiparticleApproach2024,ishizakiUnifiedTreatmentQuantum2009,tanimuraTimeEvolutionQuantum1989,tangExtendedHierarchyEquation2015,xuTamingQuantumNoise2022}
\begin{equation}\label{eq:bcf_sum}
    C(t)\approx\sum_jc_je^{i\Omega_jt},
\end{equation}
with complex prefactors $c_j$ and complex $\Omega_j=\omega_j+i\gamma_j$, whose real and imaginary parts are related to the central frequency and broadening of the peak in spectral density. \note{This requirement prevents us from directly analyzing the SD with exponential cutoff.} 
Nonetheless, we emphasize that our conclusions about the impact of the high-frequency tails of the spectral density extends beyond the HEOM method with implications for quantum analog simulation as well as other numerically exact methods that employ alternative high-frequency cutoff strategies\note{, including the exponential cutoff}.
In what follows, we will analyze the effect of using different functional forms to describe the features (i.e., peaks) of a given SD. 


The paper is organized as follows. In Sec.~\ref{the-heom} we describe the Hamiltonian model and summarize the HEOM method.  In Sec.~\ref{the-J}, we discuss representations of the spectral density peaks and, in Sec.~\ref{the-model}, we detail the model system parameters. Then, in Sec.~\ref{res}, we analyze the impact of the SD tails on the dephasing (Sec.~\ref{dephasing}) and population relaxation (Sec.~\ref{population}) of an illustrative model system (electronic excitation of a molecule in water) and suggest the way to connect simulations with different SD basis functions (Sec.~\ref{Lindblad}). In Sec.~\ref{imply} we discuss the implications and limitations of our findings.   We summarize our observations in Sec.~\ref{stn:conclusions}.

\section{\label{theory}Methods}
 We consider a quantum system coupled to harmonic bath and described by the total Hamiltonian split into the system $H_s$, bath $H_b$ and the system-bath interaction $H_{sb}=S\otimes B$ parts,
    \begin{equation}\label{eq:H}
        \hat{H} = \hat{H}_s + \hat{H}_b + \hat{S}\otimes \hat{B} =
        \frac{\hbar\Omega}{2}\hat{\sigma}_z + \sum_j \hbar\omega_j \left(\hat{a}_j^\dagger\hat{a}_j + \frac{1}{2}\right) + (\alpha_z\hat{\sigma}_z + \alpha_x\hat{\sigma}_x) \otimes \sum_j\left(g_j\hat{a}^\dagger_j+g_j^*\hat{a}_j\right),
    \end{equation}     
     where $\hat{\sigma}_z=\ket{1}\bra{1}-\ket{0}\bra{0}$ and $\hat{\sigma}_x=\ket{0}\bra{1}+\ket{1}\bra{0}$ are the usual Pauli matrices with $\ket{0}$ and $\ket{1}$ denoting the ground and excited electronic state in the bra-ket notation. The system is characterized by the transition frequency between the ground and excited electronic states $\Omega$. The bath is assumed to be a collection of harmonic oscillators with frequencies $\omega_j$, so that $\hat{a}^\dagger_j$ and $\hat{a}_j$ are the usual bosonic creation and annihilation operators. Note that the dagger symbol is used throughout to denote the adjoint of an operator.  Finally, the system-bath interaction term couples the collective bath coordinate (displacement-like bath operator) to the non-diagonal system operator weighted by $\alpha_x$ and $\alpha_z$, $\alpha_z^2+\alpha_x^2=\frac{1}{4}$.
 
In \eq{eq:H} we consider the two-level system for simplicity, but the multi-state generalization is straightforward and the conclusions of our analysis carry over. Note that the assumption of harmonic bath does not restrict our considerations to the chemical environments that are harmonic, since the effects of an arbitrary anharmonic environment can be taken into account within \eq{eq:H} by using an appropriate spectral density, provided the interaction is described well without going beyond second order in perturbation theory,\cite{leggettDynamicsDissipativeTwostate1987} which is expected for the macroscopically large environment in the thermodynamic limit, where interaction is distributed over many degrees of freedom.\cite{caldeiraQuantumTunnellingDissipative1983}

\subsection{\label{the-heom} HEOM}
We describe the dynamics of the system with the HEOM.\cite{chenBexcitonicsQuasiparticleApproach2024}
Initially the system is assumed to be in a separable state with the bath at inverse temperature $\beta$, such that the total density matrix at time 0 is
\begin{equation}\label{eq:rho0}
    \rho(0) = \rho_s(0)\otimes\rho_b^{\beta} = \rho_s(0)\otimes e^{-\beta\hat{H}_b}/Z_b,
\end{equation}
where $Z_b=\Tr_b{e^{-\beta\hat{H}_b}}$ is the bath partition function and $\Tr_b$ denotes a partial trace, taken over all bath degrees of freedom. 
To simplify the notation, we omit the hats over density operators and reserve $\rho$ with appropriate subscripts to denote a (possibly reduced) density matrix throughout.

The total dynamics of the system and environment is unitary and is generated by \eq{eq:H}. However, if only the system dynamics is of interest, the environment degrees of freedom can be traced out to yield the reduced density matrix
\begin{equation}\label{eq:rhored}
    \rho_s(t)=\Tr_b\rho(t).
\end{equation}
This reduced density matrix has nonunitary dynamics given by\cite{ishizakiUnifiedTreatmentQuantum2009}
\begin{equation}\label{eq:rhot}
    \tilde{\rho}_s(t)=\altmathcal{T}\altmathcal{F}(t,0)\tilde{\rho}_s(0),
\end{equation}
where tilde denotes that the density operator is written in the interaction picture of $H_0=H_s+H_b$, such that $\tilde{O}(t)=e^{iH_0t}O(t)e^{-iH_0t}$. Here $\altmathcal{T}$ is the time-ordering superoperator and $\altmathcal{F}$ connects the system operator $\tilde{S}$ from \eq{eq:H} but in the interaction picture to the bath correlation function from \eq{eq:bcf}
\begin{equation}\label{eq:F}
    \altmathcal{F}(t,0)=\exp\left[-\int_0^tds\tilde{S}^\times(s)\int_o^sdu\left(C(s-u)\tilde{S}(u)\right)^\times\right].
\end{equation}
The $\times$ symbol used in the superscript of the operators is the shorthand notation defined as follows:
\begin{equation}\label{eq:x}
    \hat{A}^\times \hat{B} \equiv \hat{A}\hat{B} - \hat{B}\hat{A}^\dagger.
\end{equation}
The bath correlation function depends on the interaction term in \eq{eq:H} (note that the bath operator is written in the interaction picture with respect to $H_0$ as well):
\begin{equation}\label{eq:Cdef}
    C(t)=\Tr_{b}\left(\tilde{B}(t)\tilde{B}(0)\rho^\beta_b\right),
\end{equation}
and determines the spectral density [\eq{eq:J}] that appropriately describes the environment's influence on the system dynamics.

\subsection{\label{the-J}Spectral density decompositions}
We consider a situation where the spectral density of the environment is known from experiment, simulation, or a combination of the two, and, moreover, this known spectral density consists of a broad low-frequency ($<300$~cm\textsuperscript{-1}) feature and a finite number of sharp peaks in the $300-4000$~cm\textsuperscript{-1} frequency range. This is a typical situation for molecules in solution, where the electronic energy levels are affected by several vibrational peaks, as well as the low-frequency collective motion of the solvent.

We therefore approximate the full spectral density as a sum of discrete peaks, each characterized by the peak position (frequency) $\omega_k$, peak width (broadening) $\gamma_k$, and peak intensity (reorganization energy) $\lambda_k$, resulting in the functional form $J_k(\omega;\omega_k,\gamma_k,\lambda_k)$.

The displaced Drude oscillator has Lorentzian shape
\begin{eqnarray}\nonumber
    J_k^{Dr}(\omega) &=& \frac{\lambda_k\gamma_k\omega}{\pi}\left(\frac{1}{(\omega-\omega_k)^2+\gamma_k^2}+\frac{1}{(\omega+\omega_k)^2+\gamma_k^2}\right)\\\label{eq:dr}
    &=& \frac{2\lambda_k\gamma_k(\omega^3+(\omega_k^2+\gamma_k^2)\omega)}{\pi\left((\omega-\omega_k)^2+\gamma_k^2\right) \left( (\omega+\omega_k)^2+\gamma_k^2\right)}.
\end{eqnarray}
For low frequency $\omega\rightarrow0$ and Taylor expanding \eq{eq:dr} around $\omega=0$ yields first order (Ohmic) behavior at low frequencies,
\begin{equation}
\label{eq:dr0}
    J_{k,0}^{Dr}(\omega) = \frac{2\lambda_k\gamma_k\omega}{\pi(\omega_k^2+\gamma_k^2)} + \altmathcal{O}(\omega^3).
\end{equation}
In turn, for high frequency the lowest order dependence from a Taylor expansion in $1/\omega$ around 0 is $\omega^{-1}$,
\begin{equation}
\label{eq:drinf}
    J_{k,\infty}^{Dr}(\omega) = \frac{2\lambda_k\gamma_k}{\pi\omega} + \altmathcal{O}(\frac{1}{\omega^3}).
\end{equation}
The sum in the first line of \eq{eq:dr} ensures the odd symmetry of the fit function. The integral of $J_k^{Dr}/\omega$ over the entire frequency range is the reorganization energy of the $k$\textsuperscript{th} peak, $\lambda_k$. Note that the ubiquitous Drude-Lorentz (DL) form $J^{DL}(\omega)=\frac{2\lambda\gamma\omega}{\pi(\omega^2+\gamma^2)}$ is a special case of the displaced Drude oscillator (\eq{eq:dr}) when the displacement $\omega_k$ is equal to 0. It corresponds to one summand in BCF [\eq{eq:bcf_sum}], which does not oscillate ($\omega_k=0$) but has a pure exponential decay with rate $\gamma_k$, a special case for which HEOM simulations are particularly efficient.

Performing a similar analysis for the underdamped Brownian oscillator (UBO) yields the functional form given in \eq{eq:br}. Note that the UBO SD has Ohmic behavior at low frequency [\eq{eq:br0}] with a coefficient that is double that of the displaced Drude oscillator [compare with \eq{eq:dr0}]. However, high-frequency tails fall off faster than in the Drude case, that is, as $\omega^{-3}$ vs $\omega^{-1}$.
\begin{eqnarray}\label{eq:br}
    J_k^{Br}(\omega) &=& \frac{4\lambda_k\gamma_k(\gamma_k^2+\omega_k^2)\omega}{\pi\left((\omega-\omega_k)^2+\gamma_k^2\right)\left((\omega+\omega_k)^2+\gamma_k^2\right)}\\\label{eq:br0}
    J_{k,0}^{Br}(\omega) &=& \frac{4\lambda_k\gamma_k\omega}{\pi(\omega_k^2+\gamma_k^2)} + \mathcal{O}(\omega^3) \approx 2J_{k,0}^{Dr}(\omega)\\\label{eq:brinf}
    J_{k,\infty}^{Br}(\omega) &=& \frac{4\lambda_k\gamma_k(\omega_k^2+\gamma_k^2)}{\pi\omega^3} + \mathcal{O}(1/\omega^5)
\end{eqnarray}
The UBO SD is physically motivated, as it corresponds to a situation where the system is directly and linearly coupled to a single (nuclear) mode of frequency $\omega_k$, which in turn is coupled to a bath of modes undergoing Brownian motion with friction $\gamma_k$.\cite{mukamelPrinciplesNonlinearOptical1995}

The recent proposal for the quantum analog simulator device\cite{kimAnalogQuantumSimulation2022} utilizes RLC circuits connected to the gate defined quantum dots to simulate a two-level system linearly coupled to a bosonic bath. Harmonic oscillators are mechanical analogs of the LC (inductor-capacitor) circuits and the resistive element "R" introduces broadening of discrete peaks to yield a spectral density of a functional form very similar to the UBO [compare \eq{eq:rlc} to \eq{eq:br}], but the low-frequency behavior is superohmic [\eq{eq:rlc0}] and the high-frequency tail falls off as slowly as the Drude peak, that is, as $\omega^{-1}$ but with double the prefactor [compare \eq{eq:rlcinf} to \eq{eq:drinf}].

\begin{eqnarray}\label{eq:rlc}
    J_k^{RLC}(\omega) &=& \frac{4\lambda_k\gamma_k\omega^3} {\pi\left((\omega - \omega_k)^2 + \gamma_k^2\right) \left((\omega + \omega_k)^2 + \gamma_k^2\right)} \\\label{eq:rlc0}
    J_{k,0}^{RLC}(\omega) &=& \frac{4\lambda_k\gamma_k\omega^3}{\pi(\omega_k^2+\gamma_k^2)^2} + \mathcal{O}(\omega^5) \\\label{eq:rlcinf}
    J_{k,\infty}^{RLC}(\omega) &=& \frac{4\lambda_k\gamma_k}{\pi\omega} + \mathcal{O}(1/\omega^3) \approx 2J_{k,\infty}^{Dr}(\omega)
\end{eqnarray}

Each of the three functional forms of the SD [Eqs.~\eqref{eq:dr}, \eqref{eq:br}, \eqref{eq:rlc}] affects the dynamics of an open quantum system by adding two summands to the BCF [\eq{eq:bcf_sum}], that decay exponentially with rate $\gamma_k$ and oscillate with frequency $\omega_k$. Taking $\lambda_k$, $\gamma_k$, and $\omega_k$ to be the same in all three cases presents a natural point of comparison between the three functional forms. In such a comparison the difference between the three SD functional forms is fully contained within BCF coefficients $c_j$ [\eq{eq:bcf_sum}].

In addition to the spectral density basis functions described above, we also tested more complicated functions that can be seen as generalizations of the UBO modes.\cite{ritschelAnalyticRepresentationsBath2014} Each of these 2-peak functions corresponds to two pairs of terms in the BCF, which oscillate at frequencies $\{\omega_{k1},\omega_{k2}\}$ respectively and exponentially decay with rates $\{\gamma_{k1},\gamma_{k2}\}$ respectively:

\begin{eqnarray}\label{eq:2p}
    J_k^{2,n}(\omega) &=& \Lambda_{n,k}\frac{\omega^n} {\prod_{i=\{1,2\}}\left((\omega - \omega_{ki})^2 + \gamma_{ki}^2\right) \left((\omega + \omega_{ki})^2 + \gamma_{ki}^2\right)} \\\label{eq:2p0}
    J_{k,0}^{2,n}(\omega) &=& \Lambda_{n,k}\frac{\omega^n}{\prod_{i=\{1,2\}}(\omega_{ki}^2+\gamma_{ki}^2)^2} + \mathcal{O}(\omega^{n+2}) \\\label{eq:2pinf}
    J_{k,\infty}^{2,n}(\omega) &=& \Lambda_{n,k}\omega^{-(8-n)} + \mathcal{O}(1/\omega^{10-n})
\end{eqnarray}
where the frequency scaling at the low- and high-frequency ends are $\omega^n$ and $\omega^{-(8-n)}$ respectively for $n\in\{1,3,5,7\}$. Here we employ the first two of the four possible basis functions of this form and refer to them as "Ohmic with 2 peaks" ($n=1$) and "Superohmic with 2 peaks" ($n=3$). The corresponding normalization constants $\Lambda_{n,k}$ are given in the SI.
These functions can have second order poles, but for the present discussion we will choose parameters that ensure the poles are simple. 

\note{Another form of the cutoff function -- exponential -- is often used in path-integral as well as master-equation based open quantum dynamics simulations, but cannot be directly addressed with HEOM since the exponential function has a pole of infinite order, which violates the requirement of \eq{eq:bcf_sum}. Therefore, we did not include the exponential cutoff function in our numerical simulations. Nonetheless, our analysis is general and has implications for simulations performed with the exponential cutoffs as we discuss in Sec.~\ref{Lindblad}}

\begin{figure}
    \includegraphics[width=0.49\textwidth]{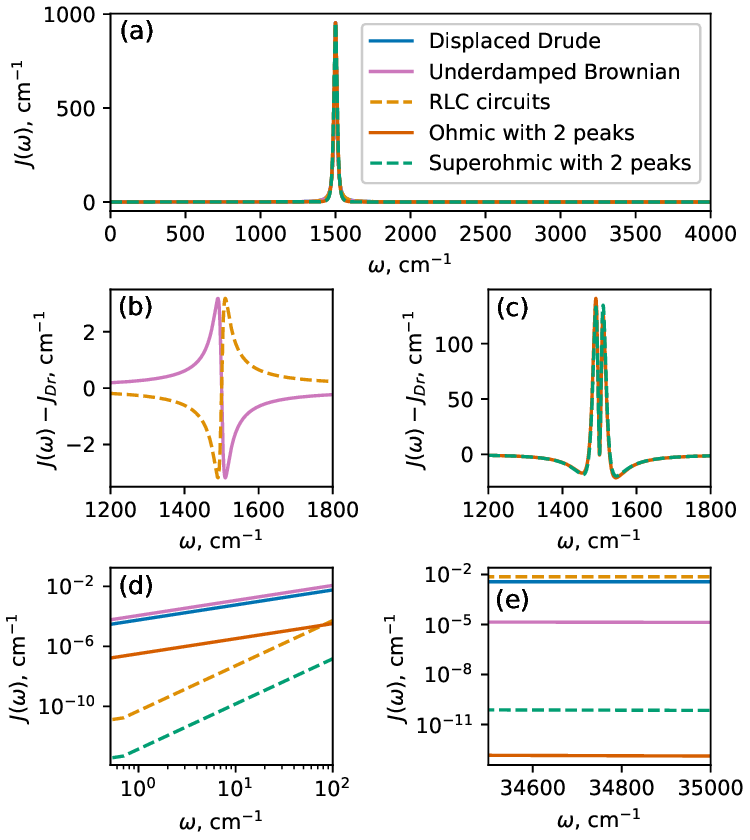}
    \caption{ \textbf{Fit functions for spectral density peaks considered in this study}. (a) A single peak at $\omega_k=1500$~cm\textsuperscript{-1} with width $\gamma_k=10$~cm\textsuperscript{-1} represented using different functional forms $J_k$. The peak intensity (reorganization energy) is set to $\lambda_k=20$~cm\textsuperscript{-1} so that the peak height is $\sim1000$~cm\textsuperscript{-1} for ease of comparison across the panels. For the two-peak functions $\omega_{k1}=\omega_{k2}=\omega_k$, $\Lambda_{2p,k}=\lambda_k$ and $\gamma_{k1}=\gamma_{k2}=2\gamma_k$. The range of frequencies shown from 0 to $4000$~cm\textsuperscript{-1} covers most chemical environments.  (b-c) difference of each peak shown in (a) with respect to the displaced Drude oscillator peak. (d) log-log plot of the low frequency tail (up to $200$~cm\textsuperscript{-1}) of each functional form shown in (a). (e) semilog plot of the high-frequency tail around $35,000$~cm\textsuperscript{-1}.
    }
    \label{fig:peaks}
\end{figure}

Rather than concentrating on the low-frequency behavior of $J_k$ (i.e., ohmic, subohmic, superohmic), we \sout{will} focus our attention on the decay of high-frequency tails at the opposite end of the spectrum. Fig.~\ref{fig:peaks}(a) shows a single peak at $1500$~cm\textsuperscript{-1} for each of the functional forms considered in this study. The five functional forms with identical peak parameters show near perfect agreement in the vicinity of the peak (Fig.~\ref{fig:peaks}b,c). The last two panels highlight the low- (Fig.~\ref{fig:peaks}d) and high- (Fig.~\ref{fig:peaks}e) frequency behavior of the different functional forms. Note that panel (d) is a log-log plot, while panel (e) has a linear x-axis and a logarithmic y-axis.

\subsection{Model system\label{the-model}}
We consider the simplest model to describe molecular decoherence in aqueous solution. The system consists of two electronic levels separated by a frequency $\Omega=35,650$~cm\textsuperscript{-1} ($4.42$~eV), corresponding to the wavelength of $280~nm$.\cite{ngErratumInitialExcitedstate2008} 
We construct the bath using the recently reported SD parameters obtained based on resonance Raman spectra \note{of thymine in aqueous solution}.\cite{gustinMappingElectronicDecoherence2023} While these parameters are obtained in the pure dephasing limit, we wish to explore the regime where bath-assisted population relaxation is also possible. We therefore couple the bath via both $\sigma_z$ with $\alpha_z=0.37$ and $\sigma_x$ with $\alpha_x=0.34$. The precise choice of coefficients is somewhat arbitrary, but to allow for both dephasing and population relaxation, they are chosen to be of comparable magnitude. We use the frequencies and reorganization energies obtained in Ref.~\citenum{gustinMappingElectronicDecoherence2023} (see Table~\ref{t:spectral_dens}). \note{We emphasize that this paper does not aim to faithfully describe the dynamics of a particular molecule (thymine), but rather to study the general phenomenon with realistic model parameters.}


Resonance Raman experiments do not inform the widths of the peaks $\gamma_k$, so we have to make a reasonable choice. The simplest option would be to pick a single width (say 10~$\mathrm{cm}^{-1}$) for all peaks in the spectral density. This works well for the displaced Drude oscillator, the underdamped Brownian, and the RLC circuit functional forms, where the intensity of each peak is controlled by the corresponding reorganization energy parameter $\lambda_k$. However, for ohmic and superohmic 2-peak functions, the reorganization energy parameter $\Lambda_{2p,k}$ sets the total peak intensity of a pair of peaks [Eqs.~(S1)-(S2)]. The peak widths in Table~\ref{t:spectral_dens} are chosen (as described in the SI) to ensure that the Ohmic 2-peak functions yield the individual peak intensities $\lambda_k$ \note{obtained from the resonance Raman}. \note{\sout{shown in Table~\ref{t:spectral_dens}}}. \note{Note that the peaks in Table~\ref{t:spectral_dens} are ordered such that neighboring peaks (e.g. 1 and 2) form a single 2-peak function.}

\begin{figure}
    \includegraphics[width=0.49\textwidth]{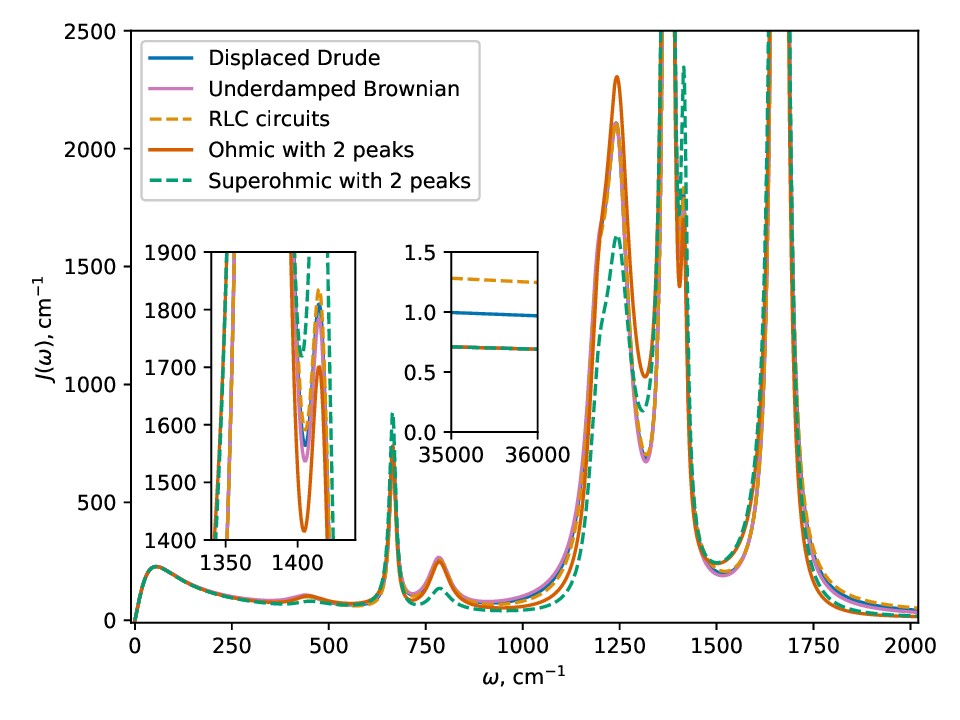}
    \caption{ \textbf{Spectral density built with different functional forms}. The parameters are shown in the table~\ref{t:spectral_dens}. The Drude, Brownian and RLC circuit functional forms result in virtually identical spectral density at the peaks with relative difference in intensity in the vicinity of the peak of less than $0.5\%$. The two-peak functions yield more pronounced differences, but overall still adequately approximate the spectral density peaks. The left inset zooms in to the range of frequencies of $1340-1440$~cm\textsuperscript{-1}, where the peaks are congested; the right inset shows the high-frequency tails in the vicinity of electronic transition frequency $\Omega=35650$~cm\textsuperscript{-1}.}
    \label{fig:J}
\end{figure}

\begin{table}
\caption{\textbf{Spectral density parameters for our model system.} The first two columns are obtained from Ref.~\citenum{gustinMappingElectronicDecoherence2023}, the third is constructed based on the behavior of Ohmic 2-peak function as described in the \sout{main text} \note{SI}.}
\label{t:spectral_dens}
\begin{tabular}{lccc}
Feature&$\omega_k$, cm\textsuperscript{-1} & $\lambda_k$, cm\textsuperscript{-1} & $\gamma_k$, cm\textsuperscript{-1}\\
\hline
Solvent & 0 & 715.7 & 54.5\\
Peak 1 & 1663 & 330 & 10\\
Peak 2 & 1243 & 161.6 & 36.55\\
Peak 3 & 1416 & 25.6 & 10\\
Peak 4 & 784 & 26.5 & 33.77\\
Peak 5 & 1376 & 186 & 10\\
Peak 6 & 1193 & 77.3 & 32.01\\
Peak 7 & 665 & 31.9 & 10\\
Peak 8 & 442 & 14.9 & 48.46\\
\end{tabular}
\end{table}
We use all five functional forms presented in Fig.~\ref{fig:peaks} to include the eight modes (peaks) of the spectral density, resulting in five similar SDs (see Fig.~\ref{fig:J}). The solvent is included via DL functional form [that is, setting $\omega_k=0$ in \eq{eq:dr}] in each of the five spectral densities. Because the low-frequency range of the spectral density is dominated by this DL solvent feature, this ensures that it is virtually identical for the five SDs we are testing. The SD peaks appear between $\sim300$~cm\textsuperscript{-1} and $\sim4000$~cm\textsuperscript{-1}. Here the SD's obtained with single-peak functional forms (displaced Drude, the UBO and the RLC circuit) display nearly perfect agreement in the vicinity of each peak, while deviating slightly more at the peak edges. The tallest peaks of the SD extend to $J(\omega_2)\approx13,000$ and $J(\omega_1)\approx18,000$ respectively;
we chose not to show the full $y$-range of the SD's, as this would make the small differences between different functional forms shown on Fig.~\ref{fig:J} indistinguishable. The two-peak functions (both Ohmic and Superohmic) visually display more pronounced deviations from the other three, but still match the peak positions, widths and intensities. The high-frequency range of the spectral density is shown in the inset and displays large relative differences between the SD's constructed with RLC circuit, the displaced Drude and the UBO peaks, as we have seen on panel~\ref{fig:peaks}(e). Note that UBO and both of the two-peak functions yield virtually identical high-frequency tails since all three functional forms decay faster than the DL solvent peak. 

The simulation is initialized in the product state $\rho(0)=\rho_s(0)\otimes\rho_b^{\beta}$, where  $\rho_s(0)=\frac{(\ket{0}+\ket{1})(\bra{0}+\bra{1})}{2}$ is the reduced density matrix of the system at the initial time and the bath is initially in the thermal state at $300$~K.
We integrate the master equation obtained based on the time-dependent variational principle using the 4\textsuperscript{th} order Runge-Kutta integrator with the 5\textsuperscript{th} order error estimator, also known as the Dormand-Prince 
algorithm.\cite{hairerSolvingOrdinaryDifferential2000} The timestep is set by the absolute and relative tolerance bounds of $10^{-8}$ and $10^{-5}$ respectively. We use the hierarchy cutoff of $25$ for each HEOM term, and add $8$ low temperature correction Pad\'{e} terms.\cite{huCommunicationPadeSpectrum2010} 

\section{\label{res}Results and Discussion}
We present the HEOM simulation results over $0.5$~ps at room temperature ($T=300$~K) in Fig.~\ref{fig:2}.

\subsection{Dephasing dynamics\label{dephasing}}
\begin{figure}
    \includegraphics[width=0.49\textwidth]{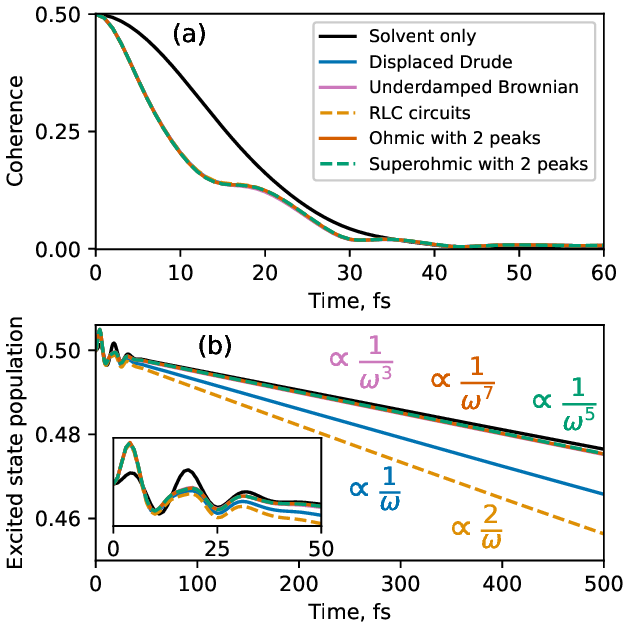}
    \caption{\textbf{Dynamics of the model system in water using different spectral density basis functions}. (a) Dephasing is insensitive to the choice of the spectral density basis functions. (b) Population relaxation is dominated by the high-frequency tails of the spectral density (SD). The color-coded rates of decay of SD peaks are shown for the different functional forms. Number 2 in the numerator of the RLC circuit scaling is to emphasize that it has a prefactor that's double that of the displaced Drude oscillators [see \eq{eq:rlcinf}]. The inset shows that even at early times (after $\sim25$~fs) the population dynamics deviates significantly depending on the choice of the functional form of SD peaks.}
    \label{fig:2}
\end{figure}
We first focus on the dephasing in the system, shown in Fig.~\ref{fig:2}a, where the absolute value of the off-diagonal element of the reduced density matrix of the system is plotted as a function of time with different functional forms of the SD peaks. The solvent alone (solid black curve in Fig.~\ref{fig:2}a) sets dephasing at $\sim30$~fs. Inclusion of the eight vibrational modes in addition to the DL solvent feature increases the initial dephasing rate, but also introduces additional structure to the dephasing dynamics. More precisely, the addition of sharp peaks in the range of frequencies between $\sim300$~cm\textsuperscript{-1} and $\sim2000$~cm\textsuperscript{-1} results in recurrences at $\sim20$~fs and $\sim35$~fs, which are not present with the solvent-only DL bath. All of these observations echo the findings presented in Ref.~\citenum{gustinMappingElectronicDecoherence2023}.

More significantly for this study, the dephasing is insensitive to the choice of the functional form of the vibrational modes in the SD, at least for the five functional forms we tested. The dephasing dynamics up to $\sim20$~fs is determined by the peak positions, widths, and intensities\cite{gustinMappingElectronicDecoherence2023} -- all of which are the same for all functional forms tested. At later times the decay of coherence is determined by the low-frequency end of the spectral density. This is because the narrow high-frequency SD features alone lead to significant coherence recurrences after the initial fast decay (Fig.~3B of Ref.~\citenum{gustinMappingElectronicDecoherence2023}) due to wavepacket evolution in alternative potential energy surfaces. In contrast, the broad low-frequency solvent feature provides an irreversible decay path for the coherence that suppresses these recurrences and dictates the overall coherence loss. Although different functional forms have different low-frequency behaviors [see Eqs.~\eqref{eq:dr0}, \eqref{eq:br0}, \eqref{eq:rlc0} and \eqref{eq:2p0}], their difference in contribution to the low-frequency end of the SD is masked by the DL solvent feature, which dominates in this frequency range (see Fig.~\ref{fig:J}). 

In summary, we find that dephasing is insensitive to the precise functional form of the structured vibrational peaks in condensed phase dynamics both at early times (controlled by molecular vibrations) and at later times (controlled by the solvent).

\subsection{Population relaxation\label{population}}

Population relaxation (Fig.~\ref{fig:2}b) is about 3 orders of magnitude slower than dephasing for our model system. This corresponds to the nonradiative relaxation time of several picoseconds -- significantly slower than dephasing, yet fast enough that the radiative mechanism (i.e., fluorescense) can be ignored.\cite{nitzanChemicalDynamicsCondensed2006}
The solvent alone causes excited state population to decay with a lifetime (inverse of exponential decay rate) of $10.5$~ps.
We observe that the rate of population relaxation is in some cases increased significantly by adding sharp peaks in the range of frequencies between $\sim300$~cm\textsuperscript{-1} and $\sim2000$~cm\textsuperscript{-1}. The key finding of this study is that this effect differs significantly depending on the functional form chosen to represent the peaks. The addition of the RLC circuit peaks yields the fastest population decay with a lifetime of $5.4$~ ps, followed by the displaced Drude oscillator peaks with a lifetime of $7.0$~ ps. Brownian and 2-peak Ohmic functions yield similar population lifetimes of $9.8$~ps, just a little faster than that caused by only the DL solvent feature. 

To explain these differences, we refer back to Fig.~\ref{fig:peaks}. The different functional forms that represent each peak in the SD have identical reorganization energies, peak widths, and characteristic frequencies, resulting in tiny absolute deviations in the vicinity of the peaks (panels a,b). However, these small differences in the functional form of the peaks significantly affect population relaxation rates. We interpret this to be caused by the large relative differences in the high-frequency tails of the five functional forms. The population relaxation rate within the Born-Markov approximation\cite{lidarLectureNotesTheory2020} is mainly determined [\eq{eq:gamma}] by the SD value in the vicinity of the frequency of electronic transition $J(\Omega)$\note{, which is significantly higher than molecular vibrations. Thus,} for the model system we study\note{, representing a typical electronic transition, the resonant contribution to electronic relaxation comes from the high-frequency tails of the SD.} 

The Brownian and 2-peak Ohmic as well as superohmic tails decay rapidly (as $\omega^{-3}$,  $\omega^{-7}$ and $\omega^{-5}$ respectively, see Eqs.~\eqref{eq:brinf} and~\eqref{eq:2pinf}). For these functional form choices the decay rate at the high-frequency end of the spectrum is higher than that of the DL solvent feature, which decays as $\omega^{-1}$, see \eq{eq:drinf}. Therefore, the presence of these peaks does not significantly influence the high-frequency tails of the overall SD (see the right inset of Fig.~\ref{fig:J}), yielding the population lifetime that is only marginally smaller than the $10.5$~ps mark dictated by the solvent alone. This result is expected from mathematical considerations, but counterintuitive, and perhaps that is why unappreciated by the quantum dynamics community. To reiterate, the DL  functional form used ubiquitously to represent the low-frequency solvent features in the spectral density not only determines the overall timescale for dephasing, but also has a dominant effect on the overall rate of population relaxation over molecular vibrations when they are represented in the SD via the UBO functional form, as is customary.

In contrast to the UBO, both the displaced Drude oscillator and the RLC circuit functional forms have peak tails that decay as $\omega^{-1}$, the same rate as the solvent feature. Therefore, vibrations represented with these functional forms can have a significant contribution to the overall SD at the high-frequency end of the spectrum. The relative importance of the solvent vs vibrations represented with either Drude or RLC functional forms depends also on the frequency of electronic transition, positions of vibrational peaks, and the difference between both reorganization energies and peak widths of the solvent vs vibrations. Note that the RLC circuits yield a faster decay rate compared to the displaced Drude oscillators because the prefactor of the dominant ($\omega^{-1}$) term at high frequency is two times larger for the former [see the inset of Fig.~\ref{fig:J} and Eqs.~\eqref{eq:drinf}, \eqref{eq:rlcinf}].

Thus, for the population relaxation calculation presented here to be accurate, the precise functional form of the vibrational peaks is needed in addition to the standard peak parameters (widths, positions, and intensities). The peak parameters can be obtained experimentally\cite{gustinMappingElectronicDecoherence2023,rengerRelationProteinDynamics2002,valleauAlternativesBathCorrelators2012} or computationally,\cite{shimAtomisticStudyLongLived2012,olbrichTheorySimulationEnvironmental2011,damjanovicExcitonsPhotosyntheticLightharvesting2002,wiethornCondonLimitCondensed2023} but extracting the precise functional form from noisy data appears unfeasible at present. Moreover, even if the functional form of $J(\omega)$ is known, its representation within the numerically exact simulation can be challenging. For example, discretization of the SD or a high-frequency cut-off can introduce error from mishandling the high-frequency end of the SD. Additionally, efficiency considerations within the HEOM calculations limit the choice of functional forms of the peaks. Finally, quantum analog simulator devices rely on  quantum hardware to construct the bath, which sets the functional form of the simulatable SD peaks. In the following section, we suggest a recipe to account for the difference between the real and the simulatable SD functional forms.

\subsection{\label{Lindblad}\note{Correcting for the difference in high-frequency tails of the SD}}
\sout{\note{Correcting simulated population relaxation rates}}
We consider a situation in which \sout{\note{the}} \textit{\note{true}} open \note{quantum} dynamics \sout{\note{of interest}} cannot be simulated due to physical or numerical constraints on the simulation, as is the case for HEOM, MCTDH, quantum analog simulation, and many other numerically exact methods. Instead, we have access to the \note{\textit{simulated}} dynamics, \sout{\note{of the \textit{source},}} which differs from the \note{true dynamics} \sout{\note{\textit{target}}} only by the functional form of the SD peaks. \note{Therefore, our goal is to recover the true population dynamics from the simulation results.}

The most straightforward way to account for the difference in population relaxation rate is to add more flexibility to the fit by increasing the number of fit functions, that is, to add more features to the simulated SD. Since the functional form of any added feature in the simulated SD does not match that of the \sout{\note{target}} \note{true SD}, we do not expect this "brute force" strategy to yield a significant improvement in either the overall quality of the fit or in the resulting population relaxation dynamics.

Another potential approach would be to perform a weighted fitting of the real SD, such that the simulated SD is more accurately fit for some frequency ranges at the cost of larger errors elsewhere. For example, in the analog simulation given in Ref.~\citenum{sunQuantumSimulationSpinboson2024} only near-resonance regions of the SD are fitted. This is justified since the effect of the environment on the dynamics of the system is not expected to be uniform across the frequency range. Such a fit could yield a better agreement in population relaxation dynamics, but it can be tricky to identify the optimal fitting weights for more complex dynamics.

Instead, we propose a simple, physically motivated, and universal approach, which can be used to adjust the results of the simulation after it has been completed. 
\note{Our }\sout{We now introduce the }key assumption\sout{, which} is that the dephasing dynamics is faster than the population relaxation dynamics by at least an order of magnitude, ensuring a separation of timescales between the two. We return with the justification of this assumption in models of realistic systems shortly. For now, we remark that this timescale separation means that the two processes are (essentially) independent. In this limit, differences in population relaxation rates due to the different functional forms in the SD peaks can be accounted for using \eq{eq:LAdjust}.

\begin{equation}
  p_{\text{real}}(t)\approx p_{eq}+\left(p_{\text{simulated}}(t)-p_{eq}\right)e^{-(\gamma_{\text{real}}-\gamma_{\text{simulated}})t}.\label{eq:LAdjust}
\end{equation}
Here $p(t)$ is the population of excited state as a function of time \note{with real and simulated SD}\sout{\note{for the (simulated) source and of the (unknown) target}}; $p_{eq}=p_\text{real}(\infty)=p_\text{simulated}(\infty)$ is the population at thermal equilibrium, and
\begin{equation}
    \gamma=2\pi \alpha_x^2J(\Omega)\coth{\left(\frac{\beta\hbar\Omega}{2}\right)}\label{eq:gamma}
\end{equation}
 is the overall population decay rate, obtained for both real and simulated SD within Born-Markov approximation [\eq{eq:gamma}].\cite{lidarLectureNotesTheory2020} This rate only depends on temperature, the frequency of electronic transition $\Omega$ and the value of the spectral density at that frequency.  \eq{eq:LAdjust} is strictly true only when the real and simulated SD's only differ by the overall population decay rate and this difference grows exponentially with time. However, we find it to be an accurate approximation as long as the population relaxation is much slower than dephasing. We derived \eq{eq:LAdjust} for population relaxation obtained via Lindblad equation (see the SI), which predicts a simple exponential decay of initial condition to approach thermal equilibrium. For our model system the Lindblad prediction fails at short times, but accurately captures the long-time dynamics and approaches the exact (thermal equilibrium) result $p_{eq}$ as time tends to infinity. This suggests that we can use \eq{eq:LAdjust} together with population relaxation rates calculated based on \eq{eq:gamma} to accurately approximate the long-time trends in population relaxation dynamics for any known SD.


This is a useful result in the context of the recently proposed quantum analog simulator device,\cite{kimAnalogQuantumSimulation2022} which uses gate-defined double quantum dots and the series of RLC circuits to simulate the system and bath parts of the spin-boson problem. As noted by the authors, the use of RLC circuits to represent the SD of the bath yields the functional form of \eq{eq:rlc}, which is similar (but not identical) to the UBO functional form [\eq{eq:br}]. The ability to establish a simple connection between the functional form obtained from RLC circuits in quantum analog simulators and other target functional forms of the SD peaks is required to simulate systems, whose spectral density peaks do not decay as $\omega^{-1}$.

\begin{figure}
    \includegraphics[width=0.49\textwidth]{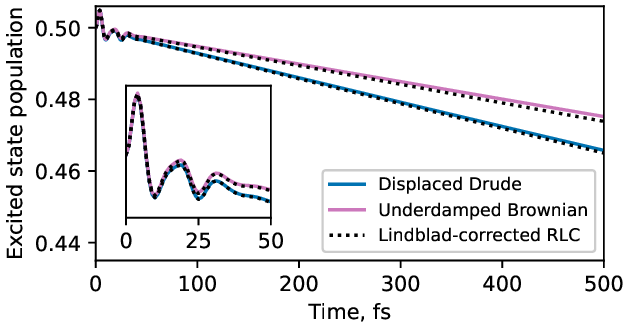}
    \caption{\textbf{RLC circuit result adjusted using \eq{eq:LAdjust}} to reproduce excited state population dynamics with displaced Drude and uderdamped Brownian functional forms of the peaks. The inset shows perfect agreement at early times (first $50$~fs).}
    \label{fig:2L}
\end{figure}

We illustrate this by taking the result of the HEOM simulation with the RLC circuit bath and adjusting it using Eqs.~\eqref{eq:LAdjust} and~\eqref{eq:gamma}. Figure~\ref{fig:2L} shows the RLC circuit results adjusted to reproduce the displaced Drude and UBO peaks as black dotted lines. The population dynamics with RLC circuit peaks differs substantially from the other two. However, upon adjustment through \eq{eq:LAdjust} we recover excellent agreement with both HEOM target results. Note that the early-time accuracy of the adjusted result does not suffer from the inability of the Lindblad equation [used to derive \eq{eq:LAdjust}] to capture early-time oscillations in the excited-state population. \note{This is because} the oscillations in the early-time dynamics  are accurately captured by either of the functional forms, leaving only the overall population decay rate to be adjusted by \eq{eq:LAdjust}.

\subsection{\label{imply}Implications and Limitations of our Findings}
In Sec.~\ref{population}, we showed and rationalized the importance of the high-frequency tails in the SD for electronic population relaxation in an instructive model system. Based on this analysis, we concluded that the high-frequency tails will dominate the nonradiative relaxation when the corresponding spectral density has no features in the vicinity of the system transition frequency. That is, when
\begin{equation}\label{eq:sep}
    \max_k(\omega_k+\gamma_k)\ll\Omega.
\end{equation}
In this regime, the precise functional form of the SD becomes important to correctly predict the relaxation dynamics. The natural question is: when do these findings become relevant to predict the relaxation behavior of molecules?

We limit our considerations to electronic transitions that preserve spin symmetry, such that there are only two competing relaxation mechanisms -- the radiative (fluorescense) and the nonradiative (internal conversion). Our analysis does not impact the description of radiative relaxation, as the spectral density for radiation fields $J(\omega)\propto \omega^3$ does not feature a cutoff frequency, so condition~\eq{eq:sep} cannot be satisfied. 

For the non-radiative mechanisms, our analysis pertains to multiphonon electronic relaxation processes in molecules. These processes proceed via phonon-mediated near-resonant transitions from the low vibrational energy levels of the excited electronic state to multiple highly excited vibrational levels of the ground electronic state. The excess vibrational energy subsequently relaxes due to rapid intramolecular vibrational energy redistribution and collisions with the solvent\cite{nitzanChemicalDynamicsCondensed2006,egorovNonradiativeRelaxationProcesses1999,rabaniClassicalApproximationNonradiative1999,egorovTheoryMultiphononRelaxation1995} These processes can be effectively modeled as a two-level system bilinearly coupled to a bosonic bath.\cite{nitzanChemicalDynamicsCondensed2006} That is, they are encoded in the bath SD and fall within the scope of our analysis.

Our considerations do not apply to electronic relaxation promoted by strong coupling to anharmonic degrees of freedom of the bath, for instance, via conical intersections.
In such cases, alternative computational approaches that can handle non-Gaussian environments are required such as the Multi-Configurational Time Dependent Hartree,\cite{meyerMulticonfigurationalTimedependentHartree1990} semiclassical methods\cite{urataniTrajectorySurfaceHopping2021} and the Automated Compression of Environments method.\cite{cygorekSimulationOpenQuantum2022} 

When constructing the model described in section~\ref{the-model} we also made several simplifying assumptions, which we do not expect to affect the validity of our conclusions. First, we truncated the electronic space to two states. Higher excited states can be added as needed, in which case the condition described by \eq{eq:sep} must be fulfilled for the dominant relaxation path. Additionally, to reduce the numerical cost, we used the spectral density extracted from pure dephasing experiments to induce both dephasing and relaxation by coupling the system to a single bath of harmonic oscillators. We do not expect the split into two independent baths to change our conclusions, since the effects of the $\sigma_x$ and $\sigma_z$ couplings for the set of parameters we have considered are essentially decoupled via separation of timescales. Furthermore, while the dissipative SD is expected to be different from that responsible for pure dephasing,  we expect that it will also satisfy \eq{eq:sep} since it originates from the same environment.

To summarise, we expect our finding to be applicable to real molecules with slow nonradiative relaxation. That is, if: (i) the spin-boson model can be used to describe the process; and (ii) electronic and vibrational energy scales are separated as described in \eq{eq:sep}. In practice, these conditions are fulfilled by many molecules that relax nonradiatively with timescales of picosecond or longer.\cite{nitzanChemicalDynamicsCondensed2006} 
In these cases, \eq{eq:LAdjust} provides a convenient way to compare \sout{\note{numerically exact}} simulations of the same system obtained by using different spectral density functions.

\section{Conclusions}
\label{stn:conclusions}

The precise structure of the high-frequency tails of the spectral density (SD) peaks is typically considered to be inconsequential for open quantum dynamics. In contrast, here we show that different peak shapes lead to identical dephasing dynamics but different relaxation rates in a spin-boson calculation when the transition frequency of the system is significantly larger than the highest frequency of the bath. 
These discrepancies arise because in this regime the population relaxation is controlled by the high-frequency tails, which are several orders of magnitude smaller than the main SD features. In such cases the full decoherence dynamics requires an accurate representation of the high-frequency tails, in addition to the main features and the low-frequency behavior of the spectral density.

These findings have several implications for the calculations of such tail-dominated population relaxation, for instance, when calculating nonradiative relaxation rates of realistic molecules. Firstly, the methods that truncate the spectral density at high frequencies will completely miss this phenomenon. Secondly, proper care must be taken to discretize the high-frequency range of the spectrum for the methods that require discretization. Thirdly, we reveal that the Drude-Lorentz solvent feature sets the overall timescale for the population relaxation if the underdamped Brownian functional form is chosen to represent vibrational peaks, as is commonly done.

We give a simple recipe for adjusting for this difference in the decay rates of high-frequency spectral density tails and discuss its application in analog simulation when the precise functional form of the target peaks is unattainable on the device. This adjustment expands the reach of such simulators to an arbitrary form of SD tails, provided
the timescales of dephasing and population relaxation are separated. Additionally, under the same requirement of timescale separation, the adjusted results can be used to directly compare simulations performed with different functional forms of the peaks. One apt example is the adjustment of an HEOM simulation performed with the (numerically advantageous) Lorentzian cutoff to the exponential cutoff, which is challenging for HEOM calculations, but common in MCTDH and other methods. 

\begin{acknowledgement}
This material is based upon work supported by the National Science Foundation under grant No. PHY-2310657. The authors thank Chang Woo Kim, John Nichol, and Ignacio Gustin for helpful discussions.
\end{acknowledgement}

\begin{suppinfo}

The supplementary information containing detailed descriptions of two-peak spectral density functions and the derivation of relaxation rate adjustments is available free of charge.

\end{suppinfo}

\bibliography{Refs}

\end{document}


\section{2-peak spectral density functions\label{app}}
The normalization constants for the two-peak spectral density basis functions are presented below. We show the simulations only for the Ohmic [\eq{eq:L41}] and the superohmic [\eq{eq:L43}] cases in Fig.~2. 
\begin{align}
    \Lambda_{1,k}&=\frac{\Lambda_{2p,k}\prod_{i=\{k_1,k_2\}}\gamma_i(\omega_{i}^2+\gamma_i^2)(\omega_i-\omega_j+\gamma_i)}{\sum_{i=\{k_1,k_2\},j\neq i}\gamma_i\left(\gamma_{i}^2 + 4\gamma_{j}^2+\omega_i^2\right)}\label{eq:L41}\\
    \Lambda_{3,k}&=\frac{\Lambda_{2p,k} \prod_{i=\{k_1,k_2\}}\gamma_i}{\sum_{i=\{k_1,k_2\}}\gamma_i}\label{eq:L43}\\
    \Lambda_{5,k}&=\frac{\Lambda_{2p,k} \prod_{i=\{k_1,k_2\}}\gamma_i}
    {\sum_{i=\{k_1,k_2\},j\neq i}\gamma_i(\omega_j^2+\gamma_j^2)}\\
    \Lambda_{7,k}&=\frac{\Lambda_{2p,k} \prod_{i=\{k_1,k_2\}}\gamma_i}
    {\sum_{i=\{k_1,k_2\},j\neq i}\gamma_{i}\left(4\gamma_i^2\gamma_j^2+4\gamma_j^2\omega_j^2+(\omega_j^2+\gamma_j^2)^2\right)}\\
    \Lambda_{2p,k}&=\frac{4 (\lambda_{k_1}+\lambda_{k_2})}{\pi}\left((\omega_{k_1}+\omega_{k_2})^2+(\gamma_{k_1}-\gamma_{k_2})^2\right)\left((\omega_{k_1}-\omega_{k_2})^2+(\gamma_{k_1}+\gamma_{k_2})^2\right)
\end{align}
To set the peak widths in Table~1 we work backwards from the functional form of the Ohmic 2-peak functions [Eq.~(20)] to ensure that the resulting spectral densities from all five basis function choices are similar with a realistic choice of widths. We pair up the eight peaks creating pairs (1,2), (3,4) etc., such that the two peaks in each pair do not have significant overlap and also have similar reorganization energy parameter ($\lambda_k$). We then set the peak widths of $10$~cm\textsuperscript{-1} for the four odd index peaks. The widths for the other four peaks (with even indices) are set such that the reorganization energies of each individual peak of the Ohmic 2-peak SD functions match the reorganization energy parameters $\lambda_k$ in Table~1.
\section{Derivation of tail adjustment\label{derive}}
We derive Eqs.~(23) and~(24) assuming both the source and target populations follow Lindblad equation.\textsuperscript{61} In this case the population of excited electronic state changes follows the rate equation:
\begin{equation}
    \dot{p}_1(t)=\gamma_-p_0(t)-\gamma_+p_1(t)\label{eq:Peom},
\end{equation}
where $\gamma_-$ and $\gamma_+$ are the (constant) rates of absorption and total (stimulated plus spontaneous) emission given in terms of spectral density as follows:
\begin{eqnarray}
    \gamma_-&=&\gamma(-\Omega)=\frac{2\pi \alpha_x^2J(\Omega)}{e^{\beta\hbar\Omega}-1}\label{eq:gamma-}=e^{-\beta\hbar\Omega}\gamma_+\\
    \gamma_+&=&\gamma(+\Omega)=\frac{2\pi \alpha_x^2J(\Omega)}{1-e^{-\beta\hbar\Omega}}\label{eq:gamma+}=2\pi \alpha_x^2J(\Omega)\left[1+\frac{1}{e^{\beta\hbar\Omega}-1}\right].
\end{eqnarray}
After elimination of $p_0(t)$ from \eq{eq:Peom} using $p_0(t)+p_1(t)=1$ (i.e., ground and excited electronic state populations add up to 1), the equation becomes
\begin{equation}
    \dot{p}_1(t)=\gamma_-[1-p_1(t)]-\gamma_+p_1(t)=\gamma_--\gamma p_1(t)\label{eq:Peom2},
\end{equation}
where $\gamma=\gamma_-+\gamma_+$.
Taking the following anzatz:
\begin{equation}
    p_1(t)=ce^{-bt}+d\label{eq:anzatz}
\end{equation}
with $c$ set by initial condition as $c=p_1(0)-d$. Plugging it into \eq{eq:Peom2}:
$$-bce^{-bt}=\gamma_--\gamma ce^{-bt}-\gamma d,$$
so $b=\gamma$ and $d=\frac{\gamma_-}{\gamma}=\frac{e^{-\beta\hbar\Omega}}{1+e^{-\beta\hbar\Omega}}=p_1(\infty)=p_1^{eq}$. Thus,
\begin{equation}
    p_1(t)=[p_1^0-p_1^{eq}]e^{-\gamma t}+p_1^{eq}\label{eq:pL},
\end{equation}
where $p_1^0=p_1(0)$ is the initial population of the excited state.

To derive Eq.~(23) we write \eq{eq:pL} for both source and target, keeping in mind that initial and final excited state populations of source and target are equal, i.e. $p_\text{target}^0=p_\text{source}^0=p_0$ and $p_\text{target}^{eq}=p_\text{source}^{eq}=p_{eq}$
\begin{eqnarray}
    p_\text{target}(t)&=&[p_\text{target}^0-p_\text{target}^{eq}]e^{-\gamma_\text{target} t}+p_\text{target}^{eq}\nonumber\\
    &=&[p_0-p_{eq}]e^{-\gamma_\text{target} t}+p_{eq}\nonumber\\
    &=&p_{\text{source}}(t)e^{-[\gamma_{\text{target}}-\gamma_{\text{source}}]t}+p_{eq}\left[1-e^{-[\gamma_{\text{target}}-\gamma_{\text{source}}]t}\right]\nonumber\\
    &=&p_{eq}+\left[p_{\text{source}}(t)-p_{eq}\right]e^{-[\gamma_{\text{target}}-\gamma_{\text{source}}]t}
\end{eqnarray}
Finally, Eq.~(24) follows from \eq{eq:gamma-} as
\begin{equation}
\gamma=\gamma_-+\gamma_+=\gamma_+\left[1+e^{-\beta\hbar\Omega}\right]=2\pi \alpha_x^2J(\Omega)\coth{\left(\frac{\beta\hbar\Omega}{2}\right)}    
\end{equation}